\documentclass[pra,twocolumn,showkeys,showpacs,superscriptaddress,floatfix,longbibliography]{revtex4-1}

\usepackage[T1]{fontenc}
\usepackage[latin9]{inputenc}
\usepackage{epsfig,psfrag}
\usepackage{latexsym}
\usepackage{graphicx}
\usepackage{dcolumn}
\usepackage{amssymb}
\usepackage{amsmath}
\usepackage{bm}
\usepackage{mathrsfs} 
\usepackage{bigints}
\usepackage{upgreek}
\usepackage{color}
\usepackage{longtable}
\usepackage{xspace}
\usepackage[squaren]{SIunits}
\usepackage{epstopdf}
\usepackage{float}
\usepackage[normalem]{ulem}
\usepackage{hyperref} 
\usepackage{bbold}
\usepackage{bbm}

\definecolor{mygreen}{RGB}{20,148,20}

\usepackage{filemod}

\newcommand{\ket}[1]{{\mid} {#1}{\rangle} }

\begin{document}

\title{Evolution of an ultracold gas in a non-Abelian gauge fields: Finite temperature effect} 

\author{Mehedi Hasan}
\affiliation{Nanyang Quantum Hub, School of Physical and Mathematical Sciences, Nanyang Technological University, 21 Nanyang Link, Singapore 637371, Singapore}
\affiliation{MajuLab, International Joint Research Unit IRL 3654, CNRS, Universit\'e C\^ote d'Azur, Sorbonne Universit\'e, National University of Singapore, Nanyang
Technological University, Singapore}
\author{Chetan Sriram Madasu}
\affiliation{Nanyang Quantum Hub, School of Physical and Mathematical Sciences, Nanyang Technological University, 21 Nanyang Link, Singapore 637371, Singapore}
\affiliation{MajuLab, International Joint Research Unit IRL 3654, CNRS, Universit\'e C\^ote d'Azur, Sorbonne Universit\'e, National University of Singapore, Nanyang
Technological University, Singapore}
\author{Ketan D. Rathod}
\affiliation{Centre for Quantum Technologies, National University of Singapore, 117543 Singapore, Singapore}
\affiliation{MajuLab, International Joint Research Unit IRL 3654, CNRS, Universit\'e C\^ote d'Azur, Sorbonne Universit\'e, National University of Singapore, Nanyang
Technological University, Singapore}
\author{Chang Chi Kwong}
\affiliation{Nanyang Quantum Hub, School of Physical and Mathematical Sciences, Nanyang Technological University, 21 Nanyang Link, Singapore 637371, Singapore}
\affiliation{MajuLab, International Joint Research Unit IRL 3654, CNRS, Universit\'e C\^ote d'Azur, Sorbonne Universit\'e, National University of Singapore, Nanyang
Technological University, Singapore}
\author{David Wilkowski}
\email{david.wilkowski@ntu.edu.sg}
\affiliation{Nanyang Quantum Hub, School of Physical and Mathematical Sciences, Nanyang Technological University, 21 Nanyang Link, Singapore 637371, Singapore}
\affiliation{MajuLab, International Joint Research Unit IRL 3654, CNRS, Universit\'e C\^ote d'Azur, Sorbonne Universit\'e, National University of Singapore, Nanyang
Technological University, Singapore}
\affiliation{Centre for Quantum Technologies, National University of Singapore, 117543 Singapore, Singapore}


\begin{abstract} 
We detail the cooling mechanisms of a Fermionic strontium-87 gas in order to study its evolution under a non-Abelian gauge field. In contrast to our previous work reported in Ref.~\cite{2201.00885}, we emphasize here on the finite temperature effect of the gas. In addition, we provide the detail characterization for the efficiency of atoms loading in the cross-dipole trap, the quantitative performance of the evaporative cooling, and the characterization of a degenerate Fermi gas using a Thomas-Fermi distribution. 
\end{abstract}

\pacs{}

\keywords{}

\maketitle 

\section{Introduction} 
Gauge fields are essential ingredients of modern theories in physics~\cite{cheng1994gauge}. By promoting the well-known $U(1)$ Abelian gauge field of quantum electrodynamics~\cite{peskin2018introduction}, to matrix valued non-Abelian gauge field has led to the understanding of electroweak interaction~\cite{Salam1959, WeinbergPRL1967, Glashow1959}, and quantum chromodynamics~\cite{marciano1978quantum}. With the advent of quantum simulation using ultracold atomic gases, it has now been possible to mimic different model-Hamiltonian from high-energy physics, condensed matter physics, astronomy~\cite{Grg2019, Barbern2015, 2012.05235, Schweizer2019, Barbiero2019,BanikExp, Eckel2018}.  

One particular thing that is common to all experiments, to perform quantum simulation of relativistic phenomena, is the competition between two specific parts of the Hamiltonian, namely the kinetic energy with quadratic dependence of momentum that has a dispersion proportional to the square-root of the temperature, and the other term arising from atom-light interaction that emulates the desired effect~\cite{lin2009synthetic,ZBSpielman2013,PhD, Spielman2020WilsonLoop}. Therefore it is imperative to lower the temperature such that the kinetic energy part of the Hamiltonian becomes sub-dominant, while the term responsible for the atom-light interaction becomes dominant~\cite{2201.00885, PhD}. 

In this paper, we focus on the realization of one specific Hamiltonian 
\begin{equation} \label{eq:Ham_exp}
    \hat{H} = \frac{\mathbf{\hat{p}}^2}{2m} - \frac{\mathbf{\hat{p}}\cdot\mathbf{\hat{A}}}{m},
\end{equation}
where $\mathbf{\hat{p}}$ is the momentum operator, $m$ is the atomic mass, and $\mathbf{\hat{A}}$ is the gauge field that is a linear combination of Pauli matrices~\cite{2201.00885, JeanDalibardRMP}. In our experiment, different components of $\mathbf{\hat{A}}$ are non-commuting, therefore the underlying gauge field is non-Abelian. The second term of the Hamiltonian in Eq.~(\ref{eq:Ham_exp}) arises from light-matter interaction and it is proportional to the single photon recoil momentum~\cite{JeanDalibardRMP}. Therefore, to observe the phenomena that arises entirely due to the second term, we need to lower the temperature so that the kinetic energy dispersion becomes sub-recoil. The aim of this present paper is to provide detail on the cooling process that has led us to simulate the Hamiltonian in Eq.~(\ref{eq:Ham_exp}), as reported in Ref.~\cite{2201.00885}. Moreover, we also provide  detail one the behaviour of the damping motion of the dynamics as a direct consequence of the weak but finite momentum dispersion of the gas.

The paper is organized as follows: In section \ref{sec:EvaporativeCooling}, we detail the evaporative cooling leading to a degenerate Fermi gas. In Section \ref{sec:Zitterbewegung}, we discuss on the observation of the dynamics of the ultracold gas in a two-dimension non-Abelian gauge field. We show that the dynamic is similar to a \textit{Zitterbewegung} effect with anisotropy of the velocity-oscillation amplitude as well as anisotropy of the oscillation damping. Finally, we conclude in Section \ref{sec:Conclusion}.

\section{Cooling of Strontium-87 to Degeneracy} \label{sec:EvaporativeCooling}
The cooling of the strontium-87 atoms in our experiment consists of two major steps, namely laser cooling and evaporative cooling. The laser cooling mechanisms have already been elaborated in Refs.~\cite{chaneliere2008three,chalony2011doppler,FDavidHighFlux}. Therefore, we detail only the evaporative cooling that allows to achieve a degenerate Fermi gas below the single photon recoil energy. 

After performing laser cooling in a magneto-optical trap (MOT), we load about $3\cdot 10^6$ atoms at a temperature of about $6\,\mu$K in a cross-dipole trap with a trap depth of $U\sim85\,\mu\mathrm{K}$. The two focused beams (waist $65\,\mu$m) are propagating in the horizontal plane, and cross at a $70^o$ angle. Their powers are controlled by acousto-optic modulators set with a $40\,$MHz frequency difference to average interference effect. The wavelength of the dipole beams is 1064 nm that is far red-detuned from the principal resonances of the strontium atom. Therefore this off-resonant dipole trap can be used to trap atoms in their ground-state at the high-intensity regions of the beams for a long time in a high vacuum environment ($\sim 50\,$s, in our case)~\cite{GRIMM200095}.

\begin{figure} 
\centering
\includegraphics[scale=0.61]{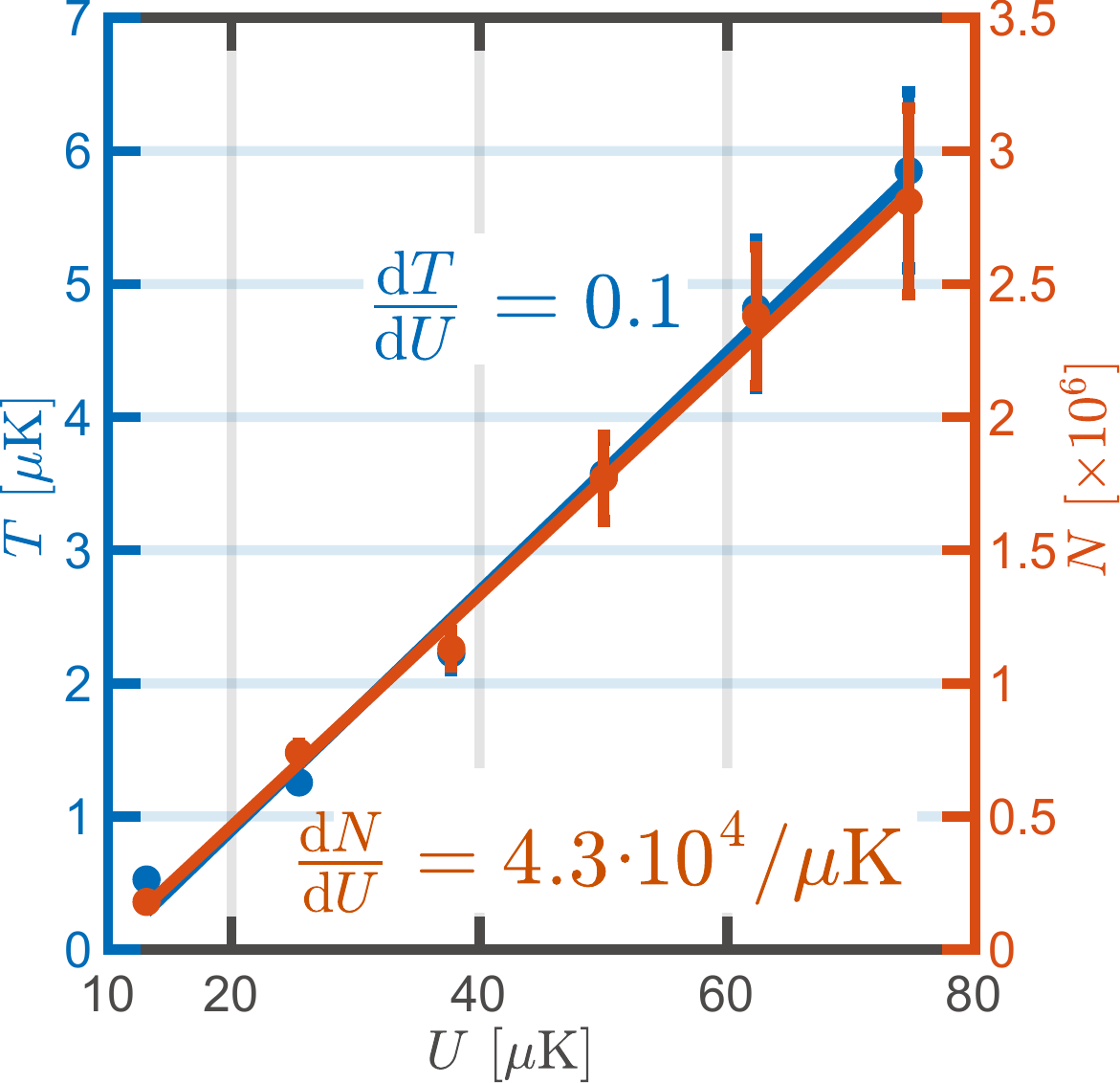}
\caption{Number of atoms loaded into the dipole trap (right axis) and the temperature of the gas (left axis), as a function of the maximum trap depth $U$. The temperature and number of atoms increases almost linearly as the $U$ is increased. The solid lines are linear fits to extract the slopes, as reported inside the main figure. The mentioned trap depth is for the ground state $^1S_0$, and is estimated theoretically from the measured power of the beams \cite{GRIMM200095}. }\label{fig:tvsTrapdepth}
\end{figure}

After loading the dipole trap, we hold the beam powers at maximal values for $3\,$s, to let the atoms thermalize and for the atoms at the wings of the each Gaussian beams to leave the trap. To characterize the loading efficiency, we quantify the number of loaded atoms and the equilibrium temperature for different values of trap depth, as shown in Fig.~\ref{fig:tvsTrapdepth}. The number of atoms increases linearly, along with a linear increase of the temperature. We see from the linear fit of the data that for every $\mu\mathrm{K}$ of trap depth $U$, we load about $4.3\cdot 10^4$ more atoms into the dipole trap at the expense of an increased temperature by 100 nK. The number of atoms in the last stage of laser cooling is about $15\cdot10^6$. Therefore at maximum trap depth, we transfer around $20\%$ of the atoms into the dipole trap.

After thermalization, we perform optical pumping to transfer all the atoms with positive $m_F$ values of $F_g=9/2$ ground states to the $\ket{F_g=9/2, m_F=9/2}$ state, while the atoms with negative $m_F$ values remain intact providing thermalization during the evaporative cooling.


\begin{figure}[b] 
\centering 
\includegraphics[scale=0.5]{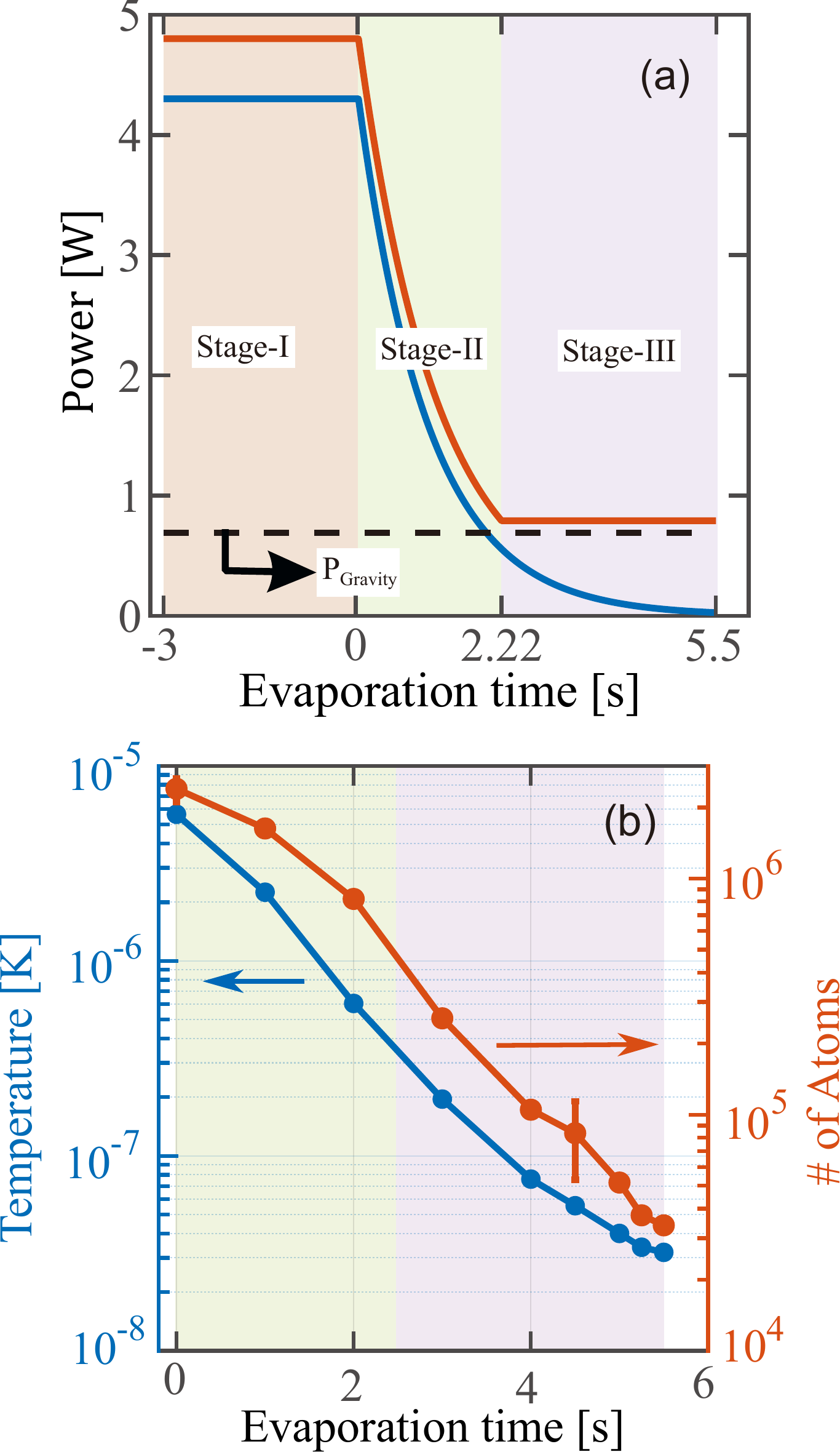}
\caption{ (a) The evaporation trajectory in our experiment with three different stages. The red and blue curves correspond to the power of the two focused beams composing our dipole trap.(b) The temperature (blue circles) and number of atoms (red circles) during the evaporation trajectory. } \label{fig:rampPower} 
\end{figure}

Our scheme for evaporative cooling is composed of three stages, see Fig.~\ref{fig:rampPower}(a). The first stage is the idle evaporation of duration $3\,$s. Afterwards, we ramp both beams, shown by blue and red lines, at a slightly different rate. This is the stage-II of the evaporative cooling, and it lasts for $2.22\,$s until one of the beams has a power that is slightly above the power necessary to hold the entire cloud against gravity, shown by the dashed black line with power $P_\mathrm{Gravity}$.  At stage-III, the power of one of the beams is held fixed while the other beam continues to lower the trap depth by decreasing the beam powers. Stage-III ends at $5.5\,$s. During Stage-III, only one beam power is lowering, therefore we call this stage 2D evaporation, in contrast to Stage-II where the trap depth is lowered in the three spatial directions namely a three dimension (3D) evaporation. \\

To quantify the full trajectory of the evaporation, we measure the number of atoms $N$ and the corresponding temperature $T$ at different time during the  evaporation, as shown in Fig.~\ref{fig:rampPower}(b). We see a continues lowering of the temperature accompanied by atom losses due to the evaporation.

To quantify the efficiency of the evaporation, we look into two particular metrics: 

1. The quantity$\frac{\partial\mathrm{Log}N}{\partial\mathrm{Log}T}$: This slope estimates the efficiency of evaporation. When this quantity is smaller than the dimensionality $D$ of the evaporation, it implies efficient evaporation, namely a net increase of the phase-space density~\cite{evaporation-molecule-jun-ye}. 

2. The quantity $T/T_\mathrm{F}$ where $T_\mathrm{F}$ is the Fermi temperature of the gas: A smaller value below unity implies a gas that going more into the degenerate regime. 

\begin{figure}[t] 
\centering
\includegraphics[scale=0.3]{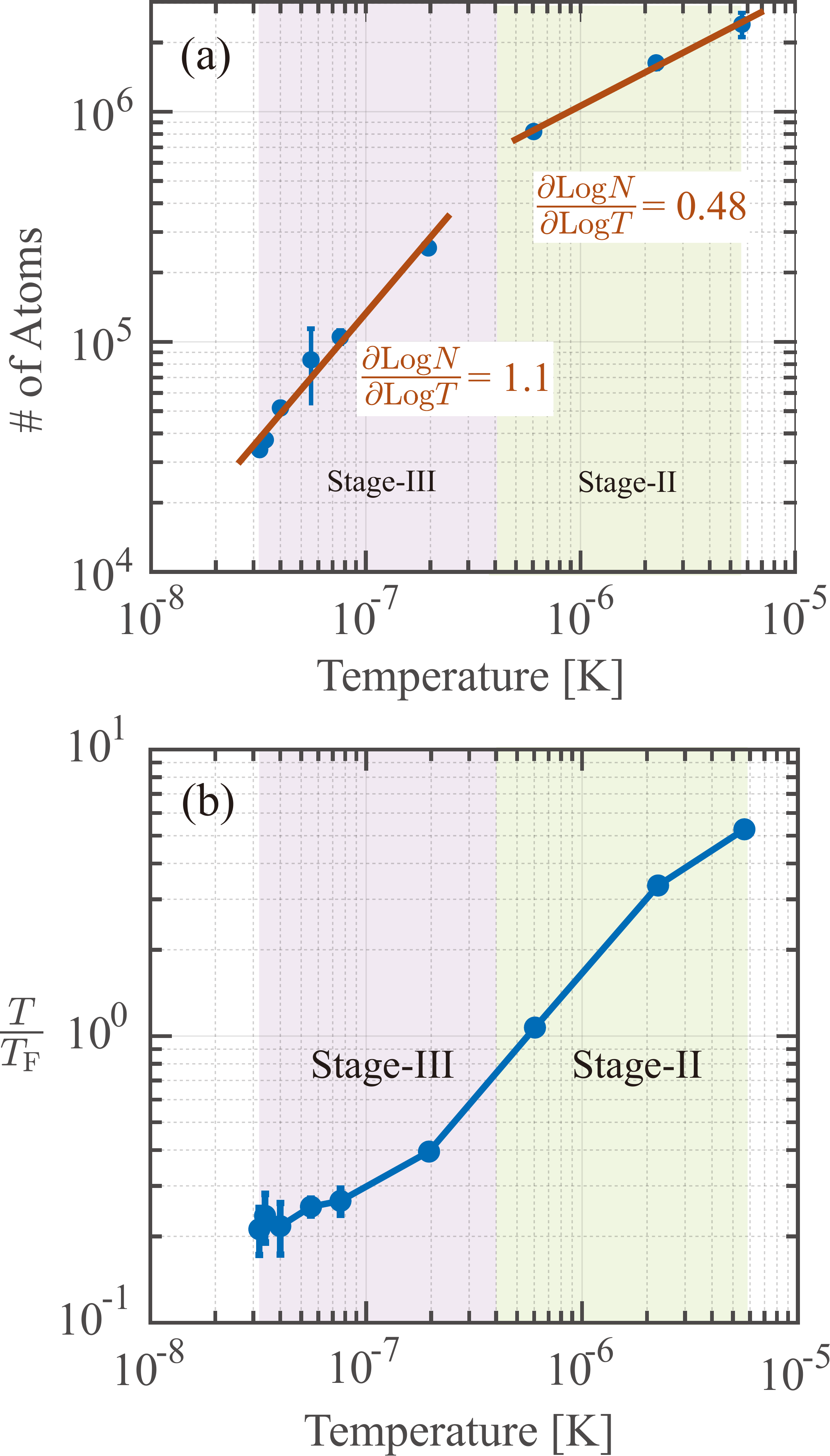}
\caption{(a) Evolution of the number of atoms as function of the temperature in the course of the evaporation. The slope $\partial\mathrm{Log}N/\partial\mathrm{Log}T<D$ is a quantitative measure of the performance of the evaporation \cite{evaporation-molecule-jun-ye}, where $D$ is the dimensionality of the evaporation. The solid lines are linear fits of the experimental data shown in blue dots. (b) Evolution of the degeneracy parameter $T/T_F$ as function of the temperature in the course of the evaporation  } \label{fig:NvsTinLogEvaporation}
\end{figure} 

These two metrics are shown in Fig.~\ref{fig:NvsTinLogEvaporation}. In  Fig.~\ref{fig:NvsTinLogEvaporation}(a), we see that during Stage-II, the evaporation is extremely efficient as the slope 0.48 in the log-log scale is much smaller than the dimensionality $D=3$. However, the efficiency drops during Stage-III. Nevertheless, the slope 1.1 is still larger than the dimensionality $D=2$. The decrease in efficiency during the later stage of evaporation is expected for a fermionic species since the Pauli exclusion principle inhibits the collisions among the atoms with same internal state, when the gas enters into the degenerate regime. The Pauli blockade is also manifested in the right panel of Fig.~\ref{fig:NvsTinLogEvaporation}, near the end of Stage-III, via the flattening of the $T/T_\mathrm{F}$. 

\begin{figure}[t] 
\centering
\includegraphics[scale=0.50]{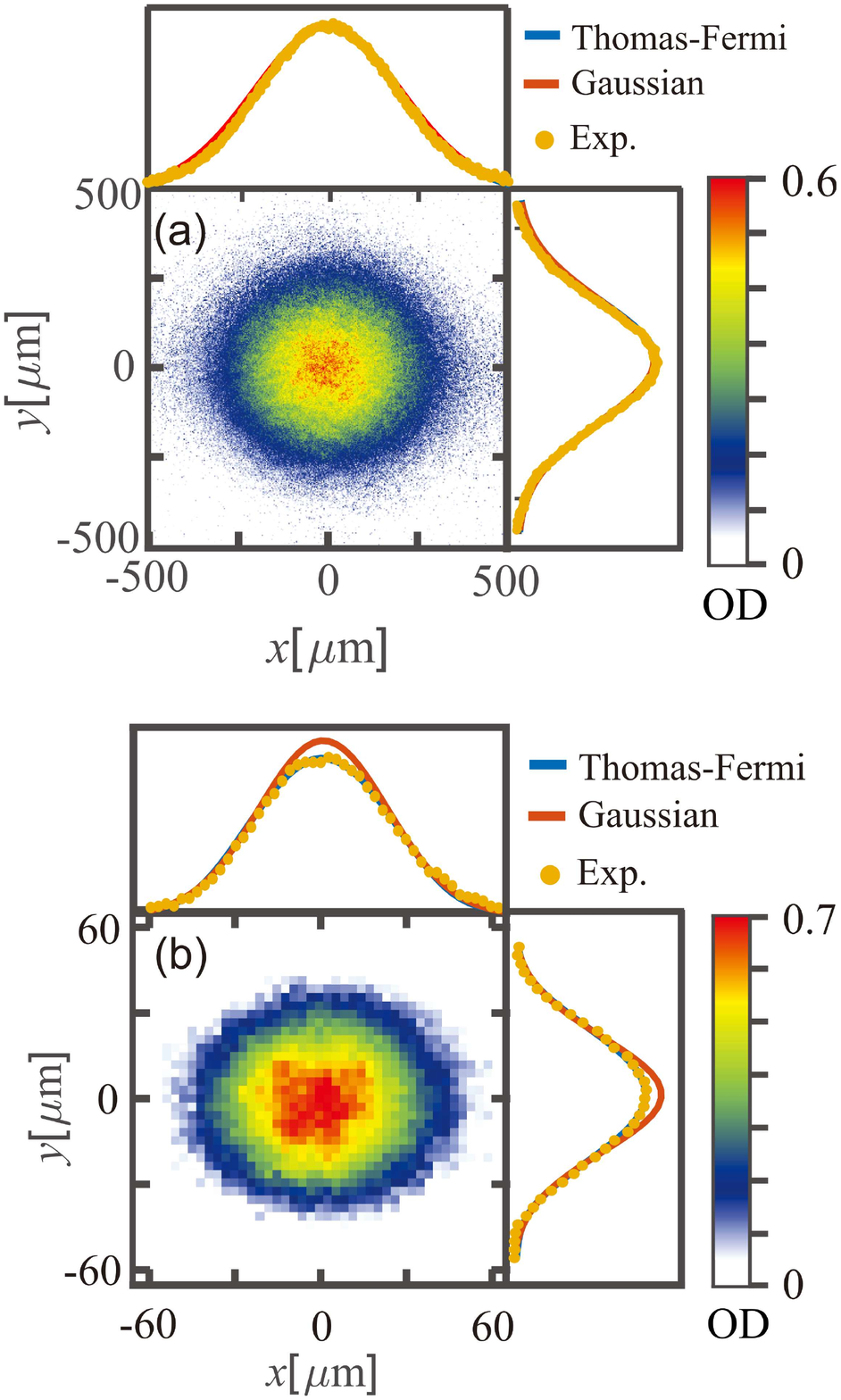}
\caption{(a) A thermal gas at $T/T_\mathrm{F}=3.1$ after 13 ms of time-of-flight. The optical density fits well with both the Gaussian (Maxwell-Boltzmann) and Thomas-Fermi distribution, indicating that the gas is essentially a classical gas. (b) Degenerate Fermi gas at $T/T_\mathrm{F}=0.21$ after 13 ms of time-of-flight. The Thomas-Fermi distribution gives the best fit while the Gaussian fit overestimates near the center of the cloud. Here, the fit is performed using the tails of the distribution, where the Maxwell-Boltzmann distribution remains a good approximation.} \label{fig:FermiGasTOF13}
\end{figure} 

At the end of evaporation, we are left with a gas of $3.7\cdot 10^4$ atoms in the state $\ket{F_g=9/2,~m_F=9/2}$ at a temperature of ${~\sim}30$ nK with $T/T_\mathrm{F}$ = 0.21. In order to extract the relevant thermodynamic quantities of the gas, we fit the momentum distribution of the gas with Thomas-Fermi distribution (see \cite{PhD} for detail). In Fig.~\ref{fig:FermiGasTOF13}(a) a thermal gas at $T/T_\mathrm{F} = 3.1$ where the momentum distribution of the gas is well described by the Gaussian function capturing a classical Maxwell-Boltzmann distribution. Note that at high values of $T/T_\mathrm{F}$, Thomas-Fermi distribution reduces to a Maxwell-Boltzmann momentum distribution~\cite{FermiFittingKetterle, PhD}. In contrast, a degenerate Fermi gas deviates from the Maxwell-Boltzmann distribution, as shown in Fig.~\ref{fig:FermiGasTOF13}b for $T/T_\mathrm{F}$ = 0.21. The essential signature of the degeneracy is the overshoot of the Maxwell-Boltzmann momentum distribution, seen via the integrated optical density along each axis in Fig.~\ref{fig:FermiGasTOF13}b. 

After evaporation, the atoms are either in $\ket{F_g=9/2,~m_F=9/2}$ or in $\ket{F_g=9/2,~m_F<0}$. The $m_F<0$ atoms are necessary to thermalize the atoms in $m_F=9/2$ during evaporation, but remain spectators for our experiment with the gauge field. As we prepare a cold gas with a temperature of $40-50\,$nK, we turn-on the appropriate laser beams to create the artificial non-Abelian gauge field for atoms in $m_F=9/2$ only and observe the evolution of the ultracold gas in the gauge field, as described in the following section.

\section{Evolution in a two-dimension non-Abelian gauge field} \label{sec:Zitterbewegung}
\begin{figure}[htb] 
\centering
\includegraphics[scale=0.12]{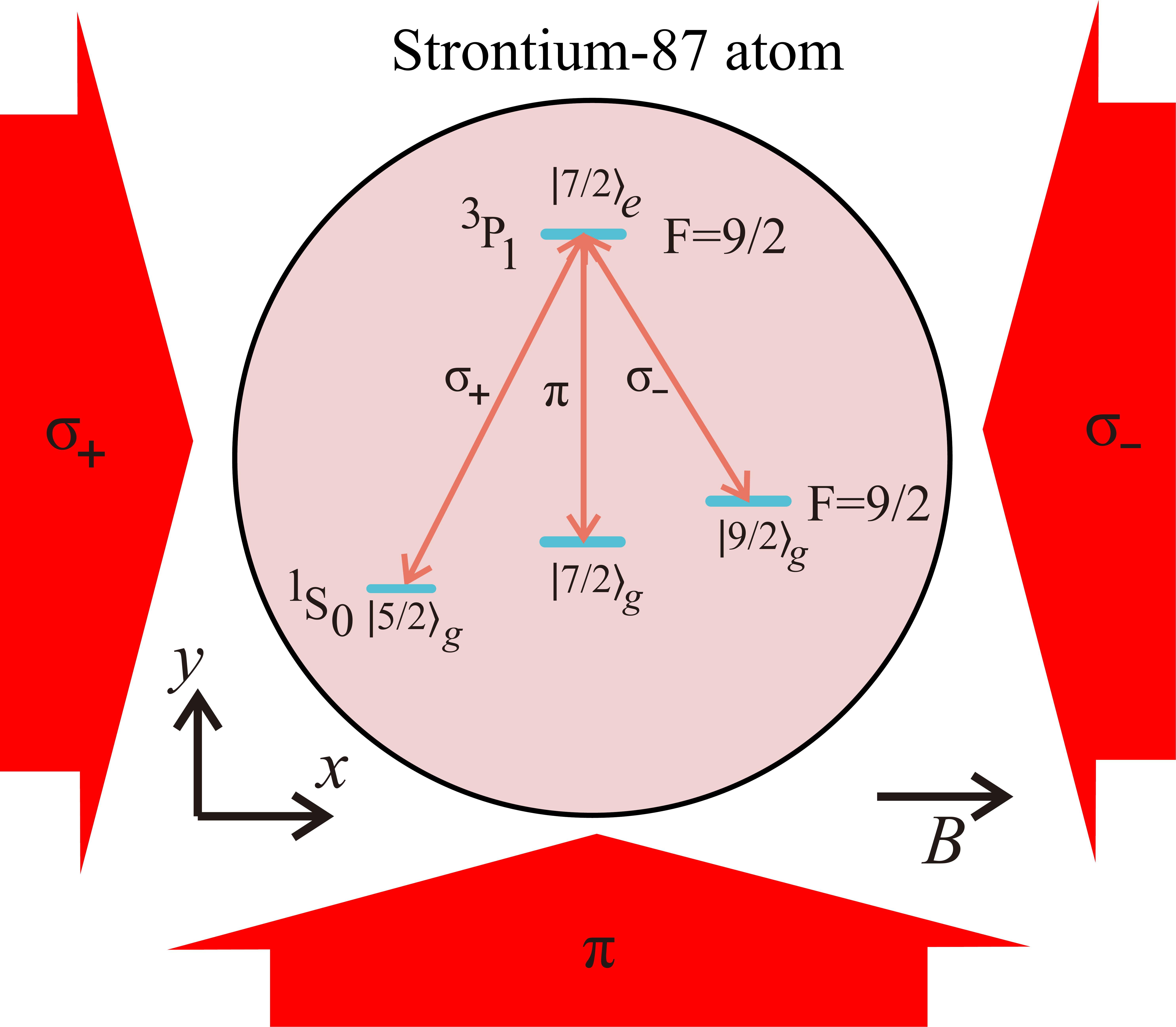} 
\caption{{Schematic of an atom showing the internal tripod scheme. The tripod is resonantly coupled with two counter-propagating laser beams along the $x$-axis with polarizations $\sigma_+$ and $\sigma_+$, and a third beam with $\pi$ polarization along the $y$-axis (red arrows). A magnetic field of $67\,$G along the $x$-axis allows to isolate the tripod scheme within the $F_g=9/2\rightarrow F_e=9/2$ hyperfine transition of the intercombination line.} }\label{fig:SchematicOfTripodScheme}
\end{figure}

We realize a 2D non-Abelian gauge field, where the governing Hamiltonian is given by Eq.~(\ref{eq:Ham_exp}). With the explicit form of the realized gauge field $\mathbf{\hat{A}}$, the matrix form of the Hamiltonian reads  
\begin{equation} \label{eq:general_H_matrix}
\hat{H} = \frac{\mathbf{\hat{p}}^2}{2m} +  \frac{\hbar k}{2\sqrt{3}m}\begin{bmatrix}
-\sqrt{3}\left(3\hat{p}_x+\hat{p}_y\right) & \left(\hat{p}_y-\hat{p}_x\right) \\
\left(\hat{p}_y-\hat{p}_x\right) & \frac{1}{\sqrt{3}}\left(3\hat{p}_x+\hat{p}_y\right), 
\end{bmatrix}
\end{equation}
where $\hat{p}_{x,y}$ are the two components of the momentum operator, and $k$ is the wave number of the laser beams that create the artificial gauge field. The Hamiltonian is expressed in the so-called dark-state basis of a tripod scheme~\cite{JeanDalibardRMP,Juzeliunas2005PRL, 2201.00885}. The latter consists of three ground states optically coupled to a unique excited state. The tripod is made of the ground states $\ket{F_g=9/2, m_F=5/2}\equiv\ket{5/2}_g$, $\ket{F_g=9/2, m_F=7/2}\equiv\ket{7/2}_g$, $\ket{F_g=9/2, m_F=9/2}\equiv\ket{9/2}_g$, whereas the excited state is $\ket{F_e=9/2, m_F=7/2}\equiv\ket{7/2}_e$, as depicted in Fig. \ref{fig:SchematicOfTripodScheme}. The two dark states, extracted from the dressed state picture, are degenerated and read,
\begin{eqnarray}
\ket{D_1} &=& \frac{e^{-2ikx} \ket{5/2}_g - e^{-ik(x+y)} \ket{7/2}_g}{\sqrt{2}}, \label{eq:D1}\\
\ket{D_2} &=& \frac{e^{-2ikx} \ket{5/2}_g + e^{-ik(x+y)} \ket{7/2}_g - 2 \ket{9/2}_g}{\sqrt{6}}, \label{eq:D2}
\end{eqnarray}
if the Rabi frequencies of the tripod beams are the same ($210~\mathrm{kHz}$ in our experiment). We note that those states do not contain the excited state and are thus immune to decoherence by spontaneous emission processes. In contrary, two other states that complete the basis are called bright states because they contain the excited state. Fortunately, they can be energetically decoupled to the dark states (adiabatic approximation). We perform our experiment under this regime, namely in the dark state manifold as discussed in Refs. \cite{FredericLerouxNatComm,2201.00885}

To transfer the atoms from $\ket{9/2}_g$ into a dark state, we turn-on the two beams with polarization $\sigma_+$ and $\pi$ shown in Fig.~\ref{fig:SchematicOfTripodScheme}. After $1~\mu\mathrm{s}$, we adiabatically turn-on the beam with polarization $\sigma_-$ that addresses the atoms in $\ket{9/2}_g$.  This scheme allows to populate the dark-state $\ket{D_2}$ of Eq. (\ref{eq:D2}). 

\begin{figure}[b] 
\centering
\includegraphics[scale=0.21]{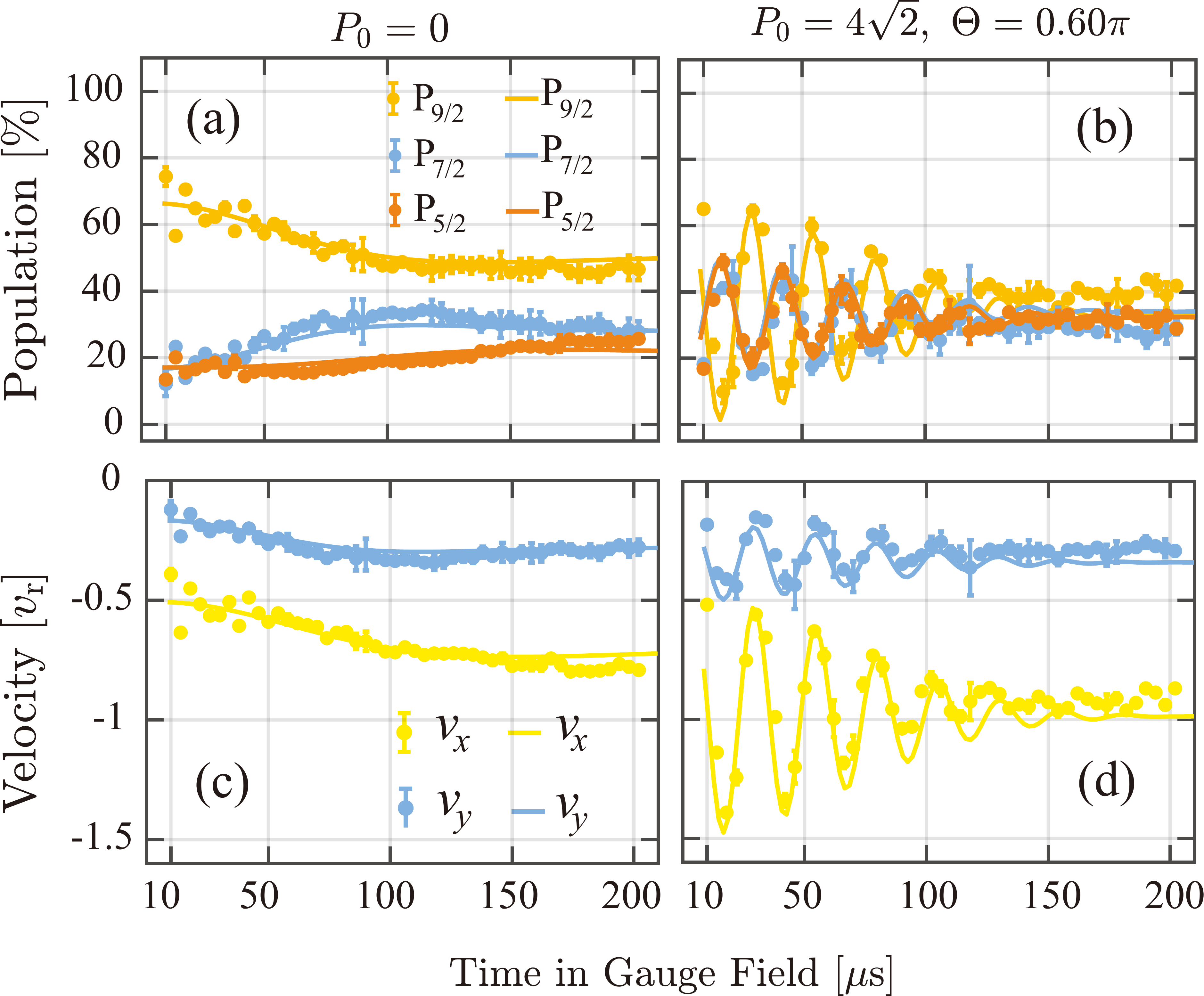}\caption{ Population and the corresponding velocity for different pushing momentum and angles. (a)-(b) the population $P_{m_F}$ of the three atomic ground states. (c)-(d) The components of velocity extracted from the population of the corresponding column. The magnitude of pushing $P_0$ is in unit of recoil momentum, and the direction is given by $\Theta$.  The dots are the extracted values of from experiments, and the solid curves are the numerical calculation using Heisenberg equation of motion without free fitting parameters~\cite{2201.00885}.  }\label{fig:PopulationAndVelocityDynamicsAtThree Situations} 
\end{figure}

After preparing the ultracold gas in the dark-state $\ket{D_2}$, we observe its dynamics at different mean momentum $\bm{P}$ with magnitude $P_0$ at an angle $\Theta$ with respect to the $x$-axis, as shown in Fig.~\ref{fig:PopulationAndVelocityDynamicsAtThree Situations}. With fluorescence imaging, we measure the population $P_{m_F}$ in the ground states $m_F$. Afterwards, we convert to the velocity along $x$- and $y$-axis via the relations \cite{2201.00885}:

\begin{equation}\label{eq:VelocityFromPopulationEqs}
{v_x} =  -v_{\mathrm{r}}(2{{P}_{5/2}} + {{P}_{7/2}}),\nonumber\\
{v_y} =  -v_{\mathrm{r}}{{P}_{7/2}},
\end{equation}
where $v_{\mathrm{r}}=\hbar k/m$ is the single photon recoil velocity.

\begin{figure}[t] 
\centering
\includegraphics[scale=0.30]{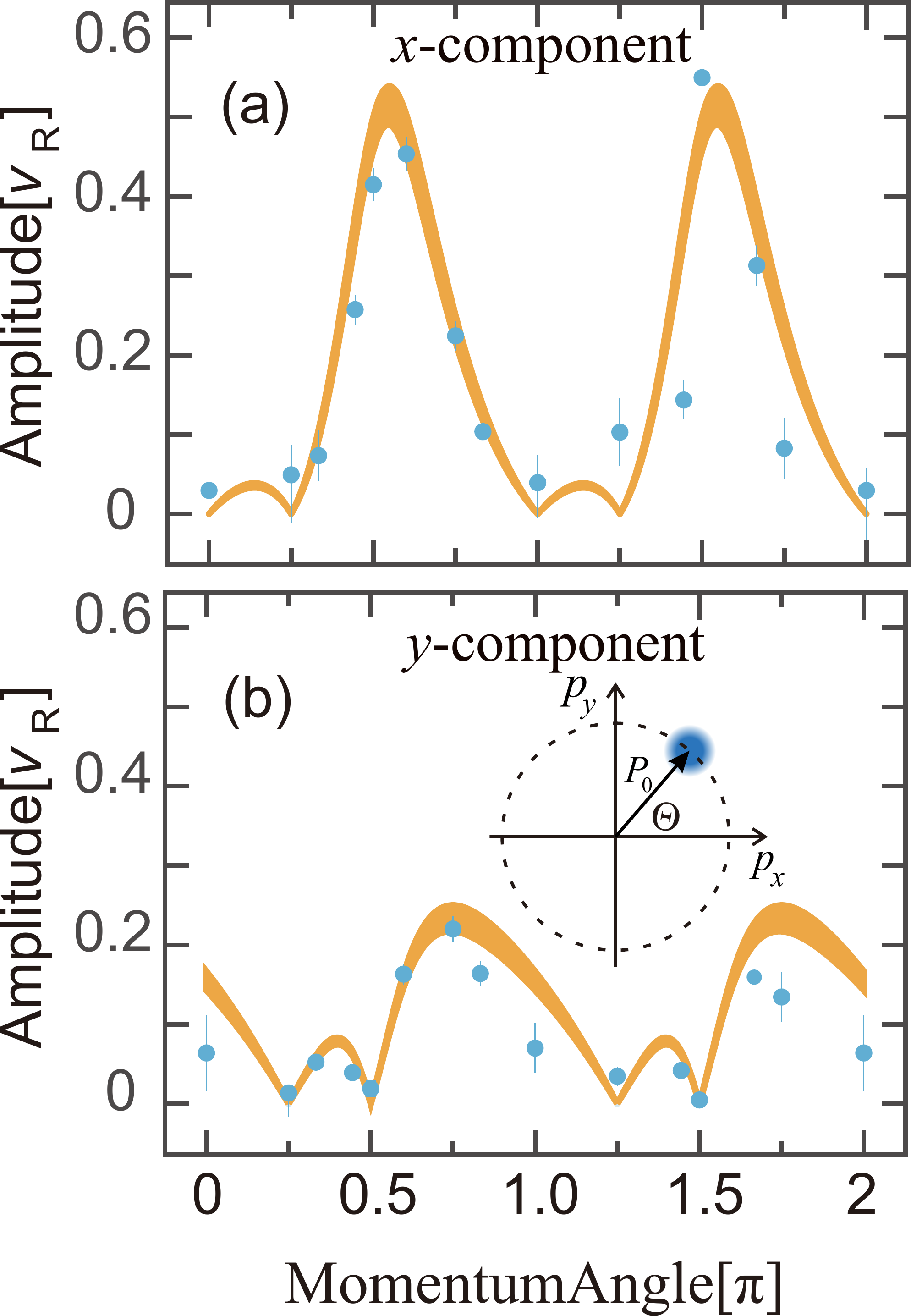}
\caption{(a) The amplitude of the $x$-component of the velocity. (b) Amplitude of the $y$-component of the velocity. The inset of (b) show the non-zero momentum of the atomic wavepacket (schematically shown as a blue disk), with magnitude $P_0=4\sqrt{2}$ at an angle $\Theta$ with the $x$-axis. Note that the solid strips on both (a) and (b) are theory prediction given by Eq.~\ref{eq:VelocityComponentsPolarForm} while accounting for finite ramping time of the tripod laser. }\label{fig:Amplitudes of both axis} 
\end{figure}

As shown in Fig.~\ref{fig:PopulationAndVelocityDynamicsAtThree Situations}, the velocity oscillation depends strongly on the mean momentum of the atoms along with the angle $\Theta$. We extract the amplitudes of both the components of the velocity as a function of $\Theta$ for a fixed value of $P_0$ and show it in Fig.~\ref{fig:Amplitudes of both axis}. Here we see that amplitudes of each axis varies strongly as a function of the $\Theta$ for a fixed value of $P_0 = 4\sqrt{2}$ given in recoil momentum unit. The expression of the velocity-components at finite temperature $T$ read \cite{2201.00885}
\begin{equation} \label{eq:VelocityExpressionsPolarFormFiniteTemperature}
    \begin{array}{l}
{v_x}\left( {{P_0},\Theta ;t} \right) \approx {v_{x1}} + {v_{x0}} \cdot \cos \left( {\omega  \cdot t} \right) \cdot \exp \left[ { - {{\left( {\frac{t}{\tau }} \right)}^2}} \right],\\
{v_y}\left( {{P_0},\Theta ;t} \right) \approx {v_{y1}} + {v_{y0}} \cdot \cos \left( {\omega  \cdot t} \right) \cdot \exp \left[ { - {{\left( {\frac{t}{\tau }} \right)}^2}} \right],\\
\\
\omega  = \frac{{2k }}{{3m}}{P_0}\sqrt {\left( {2 + \cos \left( {2\Theta } \right)} \right)} ,\\ \\ \tau  = \frac{3}{k }\sqrt {\frac{m}{{2{k_{\rm{B}}}T}}} \sqrt {\frac{{2 + \cos \left( {2\Theta } \right)}}{{5 + 4\cos \left( {2\Theta } \right)}}}, 
\end{array} 
\end{equation}

where the momentum distribution of the gas is approximated by a Maxwell-Boltzmann distribution. The offset and the amplitude of each velocity component reads:
\begin{equation} \label{eq:VelocityComponentsPolarForm}
    \begin{array}{l}
{v_{x1}} = \frac{{\hbar k }}{m} \cdot \frac{{\sin \left( {2\Theta } \right) - \cos \left( {2\Theta } \right) - 5}}{{4\left( {2 + \cos \left( {2\Theta } \right)} \right)}},\\ \\{v_{x0}} = \frac{{\hbar k }}{m} \cdot \frac{1}{{\sqrt 2 }} \cdot \frac{{\sin \left( \Theta  \right)\sin \left( {\Theta  - \frac{\pi }{4}} \right)}}{{\left( {2 + \cos \left( {2\Theta } \right)} \right)}};\\
\\
{v_{y1}} = \frac{{\hbar k }}{m} \cdot \frac{{3\sin \left( {2\Theta } \right) - 5\cos \left( {2\Theta } \right) - 7}}{{12\left( {2 + \cos \left( {2\Theta } \right)} \right)}},\\ \\ {v_{y0}} = \frac{{\hbar k }}{m} \cdot \frac{1}{{\sqrt 2 }} \cdot \frac{{\cos \left( \Theta  \right)\sin \left( {\frac{\pi }{4} - \Theta } \right)}}{{\left( {2 + \cos \left( {2\Theta } \right)} \right)}}.
\end{array}
\end{equation} 

The frequency of the velocity oscillation $\omega$ in Eq.~(\ref{eq:VelocityExpressionsPolarFormFiniteTemperature}) depends both on the magnitude and the angle of $\bm{P}$ while the amplitudes of the motion given by the expressions in Eq.~(\ref{eq:VelocityComponentsPolarForm}) depend only on the angle $\Theta$. This $\Theta$-dependence leads to the anisotropic behaviour observed in the experiment. Note that both velocity components vanish at $\Theta=\pi/4$ and $5\pi/4$. At these angles, the dark-state $\ket{D_2}$, which is the initial state for all the experiments, is an eigenstate of the Hamiltonian, and this leads to the suppression of the oscillation ~\cite{2201.00885}. 

\begin{figure}[t] 
\centering 
\includegraphics[scale=0.46]{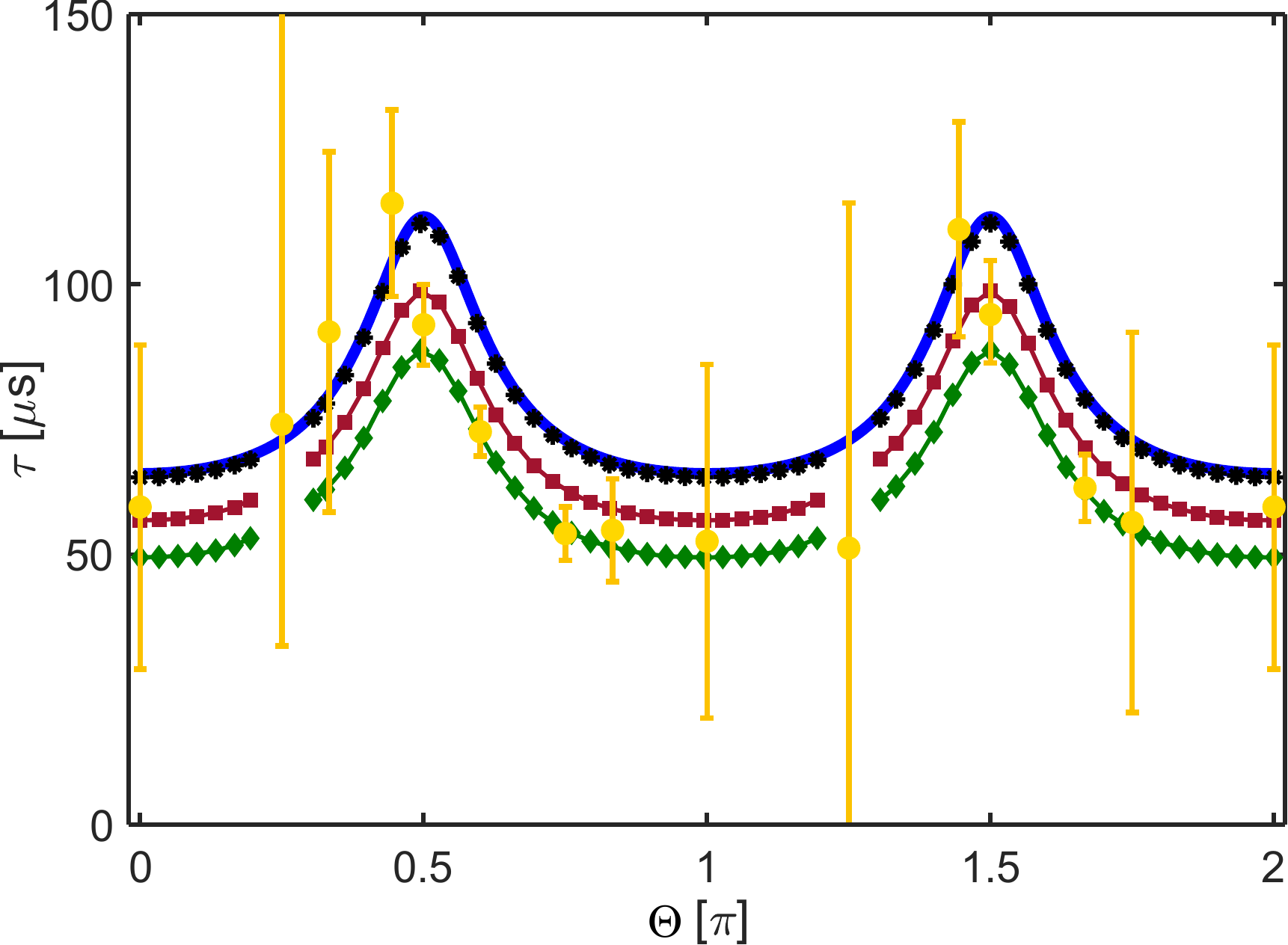}
\caption{Damping time $\tau$ of the velocity. The orange circles are the result of the fitting with an damped oscillation of the experimental data. The plain blue curve is the theory prediction from Eq.~(\ref{eq:VelocityExpressionsPolarFormFiniteTemperature}). The black stars, red squares, and green diamonds correspond to a simulation with a Maxwell-Boltzmann, a Fermi-Dirac distribution with $T/T_F=0.3$, and  a Fermi-Dirac distribution with $T/T_F=0.2$, respectively. The fitting procedure is disregarded at the vicinity of $\Theta=\pi/4$ and $\Theta=5\pi/4$ where the oscillation vanishes.}\label{fig:Damping of both axis} 
\end{figure} 

The anisotropy of the damping time $\tau$ of the velocity oscillation is shown in Figs.~\ref{fig:Damping of both axis}. Here we see that the damping time $\tau$ (Orange circles) depends on the direction of the mean momentum imparted on the atoms. The solid blue curve is the theory prediction of $\tau$ from Eq.~(\ref{eq:VelocityExpressionsPolarFormFiniteTemperature}) where we have used a Maxwell-Boltzmann velocity distribution. Note that although the theory captures the trend of the experimental data, the estimation from theory generally overestimates the damping time measured in the experiment. We also performed a simulation of the dynamic of the gas using the Hamiltonian \ref{eq:Ham_exp} in the Heisenberg approach. We then fit the damping of the oscillation using an exponential decay. With a Maxwell-Boltzmann distribution (black stars), the fitted decay time is in perfect agreement with the theory prediction from Eq.~(\ref{eq:VelocityExpressionsPolarFormFiniteTemperature}). We also take into account the role of the Fermi statistic, with decays obtained for $T/T_F=0.3$ ($T/T_F=0.2$) corresponding to the red squares (green diamonds). The reduction of the damping time with a Fermi-Dirac distribution can be simply understood by a reduction of the low momentum population with respect to the Maxwell-Boltzmann distribution.

\section{Conclusions} \label{sec:Conclusion}
In conclusion, we described the loading of atoms in a cross-dipole trap and evaporative cooling to reach a degenerate Fermi gas with $T/T_\mathrm{F} = 0.21$. The evaporative cooling of the atoms with proper characterization of the efficiency is described in detail. This cold gas is used further to observe the anisotropic \textit{Zitterbewegung} dynamics of the ultracold gas in a 2D non-Abelian gauge field.  The role of Fermi degeneracy is emphasized to understand extra damping of the velocity dynamics.

\textit{Acknowledgments:} This work was supported by the CQT/MoE funding Grant No. R-710-002-016-271, and the Singapore Ministry of Education Academic Research Fund Tier1 Grant No. MOE2018-T1-001-027 and Tier2 Grant No. MOE-T2EP50220-0008.\\

\bibliography{ms}
\end{document}